\documentclass[9pt]{elife}

\usepackage{lipsum} 
\usepackage[version=4]{mhchem}
\usepackage{siunitx}
\usepackage{cleveref}
\DeclareSIUnit\Molar{M}
\usepackage{natbib}
\usepackage{changepage}
\usepackage{nicefrac}
\usepackage{bm}
\renewcommand{\b}{\bm}
\usepackage{placeins}

\title{Metabolic Model-based Ecological Modeling for Probiotic Design}

\author[1,2*]{James Brunner}
\author[3,4*]{Nicholas Chia}
\affil[1]{Biosciences Division, Los Alamos National Laboratory, Los Alamos, NM, USA}
\affil[2]{Center for Nonlinear Studies, Los Alamos National Laboratory, Los Alamos, NM, USA}
\affil[3]{Microbiome Program, Center for Individualized Medicine, Mayo Clinic, Rochester, MN, USA}
\affil[4]{Department of Surgery, Mayo Clinic, Rochester, MN, USA}

\corr{jdbrunner@lanl.gov}{JB}
\corr{chia.nicholas@mayo.edu}{NC}

\begin{document}

\maketitle

\begin{abstract}
The microbial community composition in the human gut has a profound effect on human health. This observation has lead to extensive use of microbiome therapies, including over-the-counter ``probiotic" treatments intended to alter the composition of the microbiome. Despite so much promise and commercial interest, the factors that contribute to the success or failure of microbiome-targeted treatments remain unclear. We investigate the biotic interactions that lead to successful engraftment of a novel bacterial strain introduced to the microbiome as in probiotic treatments. We use pairwise genome-scale metabolic modeling with a generalized resource allocation constraint to build a network of interactions between 818 species with well developed models available in the AGORA database. We create induced sub-graphs using the taxa present in samples from three experimental engraftment studies and assess the likelihood of invader engraftment based on network structure. To do so, we use a set of dynamical models designed to reflect connect network topology to growth dynamics. We show that a generalized Lotka-Volterra model
has strong ability to predict if a particular invader or probiotic will successfully engraft into an individual's microbiome.
Furthermore, we show that the mechanistic nature of the model is useful for 
revealing which microbe-microbe interactions potentially drive engraftment.
\end{abstract}

\section{Introduction}
Microbiome research has come to encompass key areas of disease, ranging from infections \citep{antharam2013intestinal,honda2012microbiome,battaglioli2018clostridioides} and cancer prevention \citep{moss2005mechanisms,walther2016potential,kim2020fecal} to systemic immune and neurological responses \citep{severance2016autoimmune,kang2014diet,chen2016multiple}. The effect of the microbiome on health is now undeniable, and every year in the US over 400,000 people collectively spend \$1 billion dollars on over-the-counter probiotics intended to alter their microbiome\citep{kristensen2016alterations}. 
Many of the purported interactions between microbes and health involve {\it resident} microbiota and their interactions with the host, i.e., the interface between microbial ecology and human health. The goal of microbiome-targeted interventions is therefore to promote health by ``restoring and maintaining the microbiota and the crucial health-associated ecosystem services that it provides" \citep{costello2012application}.

Despite the many links between the microbiome and health, our ability to deploy probiotics to modify the microbiome as intended has been met with relatively little success\citep{mullard2016leading,zhu2019bifidobacterium,yuan2017efficacy,zhao2021meta,wu2017effect}. Studies looking at the ecological effects of probiotic administration show that administration of a probiotic is not sufficient to alter the community in the desired way. Specifically, engraftment of the administered microbial species is often limited, with only one-third to one-half of patients showing any signs of medium or long-term engraftment\citep{maldonado2016stable,pudgar2021probiotic}. We offer the argument that probiotic interventions are primarily ecologic in nature; their purpose is to reshape the complex microbial communities in our body in beneficial ways. Therefore, to predict whether a probiotic has the desired effects in the gut microbial community, we need more studies examining the ecology of probiotic interventions\citep{walter2018engraft}. A mechanistic, personalized approach to probiotic design---one rooted in empirical metabolic data and ecologic principles---has the potential to propel the field forward. 



Previous work trying to predict engraftment has been mostly restricted to fecal microbiome transplant (FMT) studies using non-mechanistic classifiers\citep{smillie2018strain,podlesny2021intraspecies}. While potentially predictive, such classifier approaches are sensitive to the underlying assumptions or conditions in which the study is carried out. Because such models are built using statistical methods on data that is assumed to be uniformly collected, these models cannot be generalized to new circumstances. It would be unexpected to see predictions built from patients with diarrhea, undergoing bowel prep, taking antibiotics, and given an FMT accurately predict what happens in patients taking orally administered probiotics. 


The goal of this work is to examine the use of metabolic modeling-informed population dynamic approaches. This builds on top of related work in both constraint-based metabolic modeling and population models such as Lotka-Volterra. It is worth highlighting that the use of dynamic flux balance analysis for population models has also been a well-published approach. Despite these successes, there are a number of practical drawbacks for communities of high complexity such as labor-intensive interpretation and high computational complexity.

\begin{figure}
    \centering
    \includegraphics[scale = 0.65]{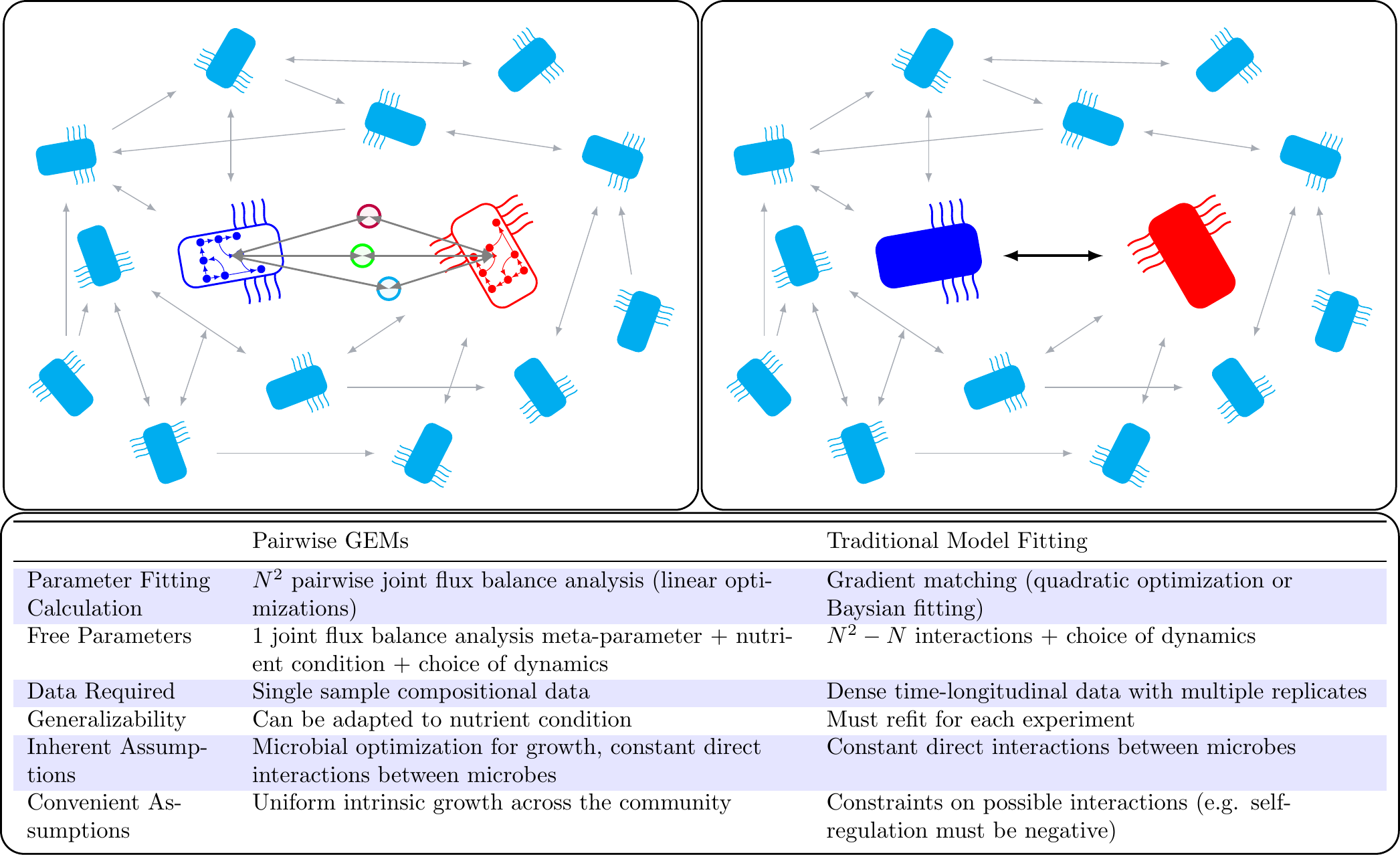}
    \caption{Population models such as Lotka-Volterra generally require dense time-longitudinal data to accurately parameterize. In this work, we leverage genome-scale metabolic modeling to parameterize population models with only genomic data from a single time-point. We are unable to parameterize these models using standard techniques due to the sparsity of the data.}
    \label{fig:improvement}
\end{figure}

Population models such as the generalized Lotka-Volterra model are popular tools for understanding microbial community dynamics in a mechanistic manner \citep{stein2013ecological,gore2017,angulo2019theoretical,kuntal2019web}. However, these models are in general difficult to parameterize, with state of the art gradient-matching procedures requiring somewhat dense time-longitudinal data with many replicates \cite{bucci2016mdsine}. Furthermore, the authors have previously shown that parameters fit from data to these models do not extend to novel environmental situations, and may even change with the addition of a new taxa to the community \citep{brunner2019metabolite}. These drawbacks make such mechanistic population models impractical for predicting engraftment. However, by leveraging metabolic modeling we are able to parameterize population models in a way that can be easily adapted to novel environments and does not require dense time-longitudinal data. This allows us to use these models to predict microbial engraftment into a community. See \cref{fig:improvement} for a comparison of our method with standard parameter fitting.

\begin{figure}
    \centering
    \includegraphics[scale = 0.65]{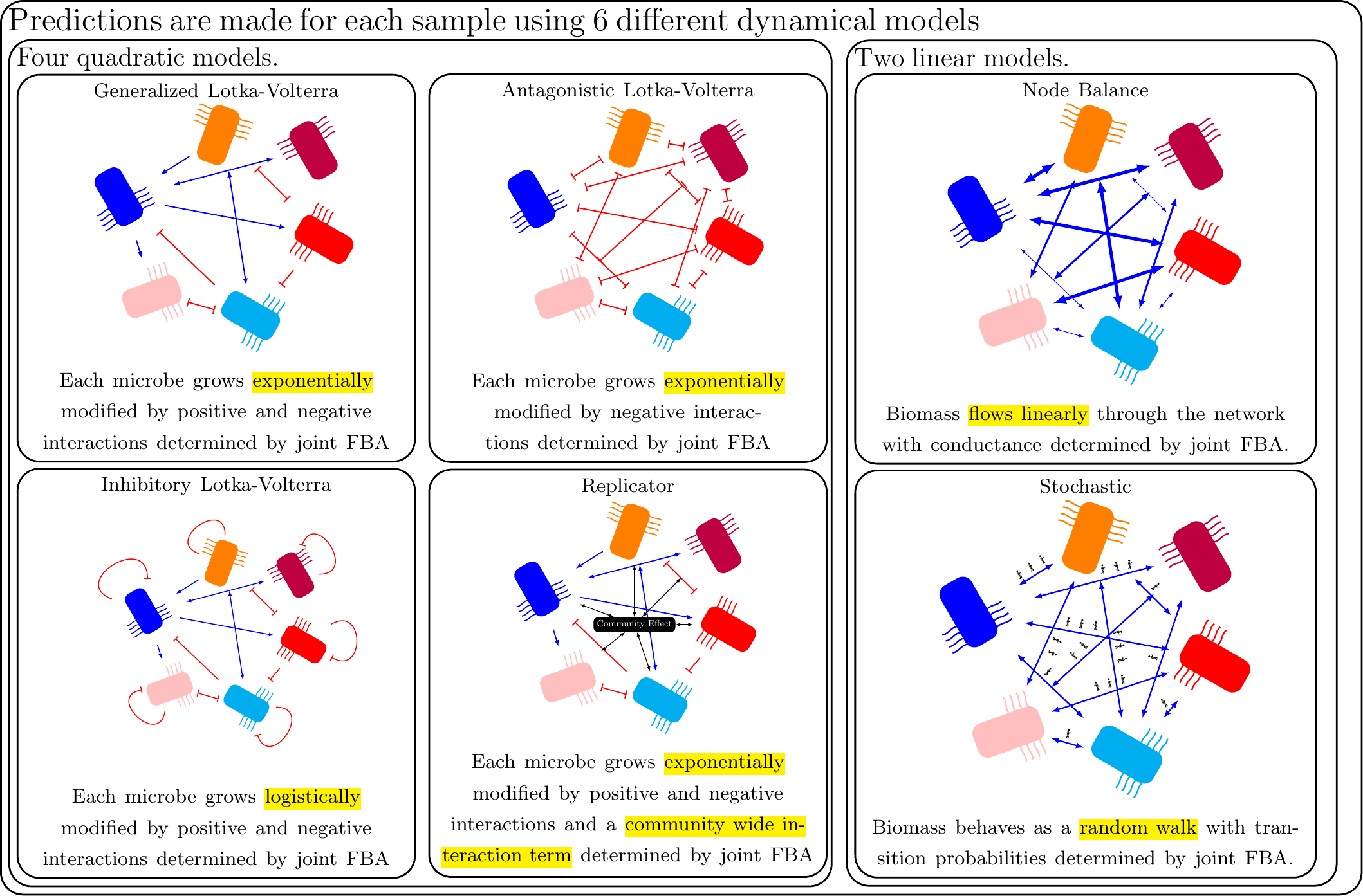}
    \caption{We used six dynamical systems to assess the engraftment potential of an invading taxa. All six are parameterized based on a set of network connections $w_{ij}$ derived from metabolic modeling. The four quadratic models are variations on the generalized Lotka-Volterra model. The two linear models are related to diffusion on the graph.}
    \label{fig:models}
\end{figure}

In this paper, we present a method to predict engraftment of an invader into a microbial community in the following manner. First, we construct an interaction network of the microbial taxa found in a sample from the community using pairwise flux balance analysis with resource allocation constraints \citep{kim2022resource} with genome-scale metabolic models included in the AGORA database\citep{magnusdottir2017generation}. We then make a prediction based off of one or more of six dynamical systems models parameterized by this network---the generalized Lotka-Volterra (LV) model \cite{edelstein2005mathematical,stein2013ecological,gore2017}, an adjustment to the LV model we call the ``antagonistic Lotka-Volterra" model (AntLV), another adjustment to the LV model we call the ``inhibitory Lotka-Volterra" model (InhibtLV), a fourth non-linear model based on the replicator equation \cite{madec2020predicting,gjini2021ratio} which is similar to the LV model, and two similar linear models based on node balancing (NodeBalance) and random walks (Stochastic) (see \cref{fig:models}). 

We test the predictive potential of each dynamical model by predicting the outcomes of microbiome invasion experiments \citep{battaglioli2018clostridioides,maldonado2016stable,huang2021candidate} from the initial presence/absence of species in each sample. We choose to test all six models because no definitive ``best model" has been established. The LV model is widely used, but we found that this model could lead to uncontrolled simulated growth. We developed the AntLV and InhibitLV models in order dampen the numerical instability of the LV model. We also include the replicator model because has recently been shown to provide insight into microbial population dynamics \citep{madec2020predicting}. Finally, we include the two linear models because they provide a substantial reduction in computational complexity, and so are more practical to use as long as they provide adequate predictive power.

Because the specific invasion experiments involved invader strains not present in the AGORA database, we repeated the experiment for each species-level match to the invader in the database. We show that this method has good predictive potential depending on the choice of dynamical system used to score the sub-graphs and choice of AGORA database match to the invader strain. Furthermore, we perform two types of sensitivity analysis to demonstrate that the mechanistic nature of the model provides additional insight into the impact of the various components of the network. We perform simulated knock-out experiments to test how sensitive engraftment is to each community member, and we use parameter sensitivity analysis to test how sensitive engraftment is to each network connection.

\section{Results}
We first examine the ability of our metabolic modeling-based approach to successfully predict engraftment versus non-engraftment for different microbial species introduced orally across different experimental or clinical trial settings. The three studies we use to test predictive power all involved the introduction of an invading taxa into an established microbial community. Two of these studies examine the introduction of candidate probiotics, \emph{B. longum} and \emph{L. plantarum}, respectively, into human subjects \citep{maldonado2016stable,huang2021candidate}. The third examines the introduction of the pathogen \emph{C. difficile} \citep{battaglioli2018clostridioides} into mice that had been inoculated with microbial communities from healthy and disbiotic human donors. Note that while this study introduces a pathogen rather than a probiotic, the ecological principle of engraftment into an existing community is the same. All three studies include microbiome samples from prior to introduction of the invader, which we use to generate predictions, as well as samples from after the community was allowed to reform, which we use to score our predictions.

As a metric of classification success for the \emph{C. difficile} and \emph{B. longum} data-sets, we use the area under the curve of the receiver-operator characteristic (AUC-ROC). This metric provides a measure of performance based on the model's ability to identify true positives while avoiding false positives, so that $1$ is perfect classifier performance and $0.5$ is equivalent to random classification (i.e. flipping a coin for each sample). We use Kendall-tau rank correlation as a metric of classifier success for the \emph{L. plantarum} data-set, comparing our test-score with the observed abundance. This is necessary because AUC-ROC requires binary classification of the test data, which was unavailable for this data-set.

\begin{figure}
    \centering
    \includegraphics[scale = 0.3]{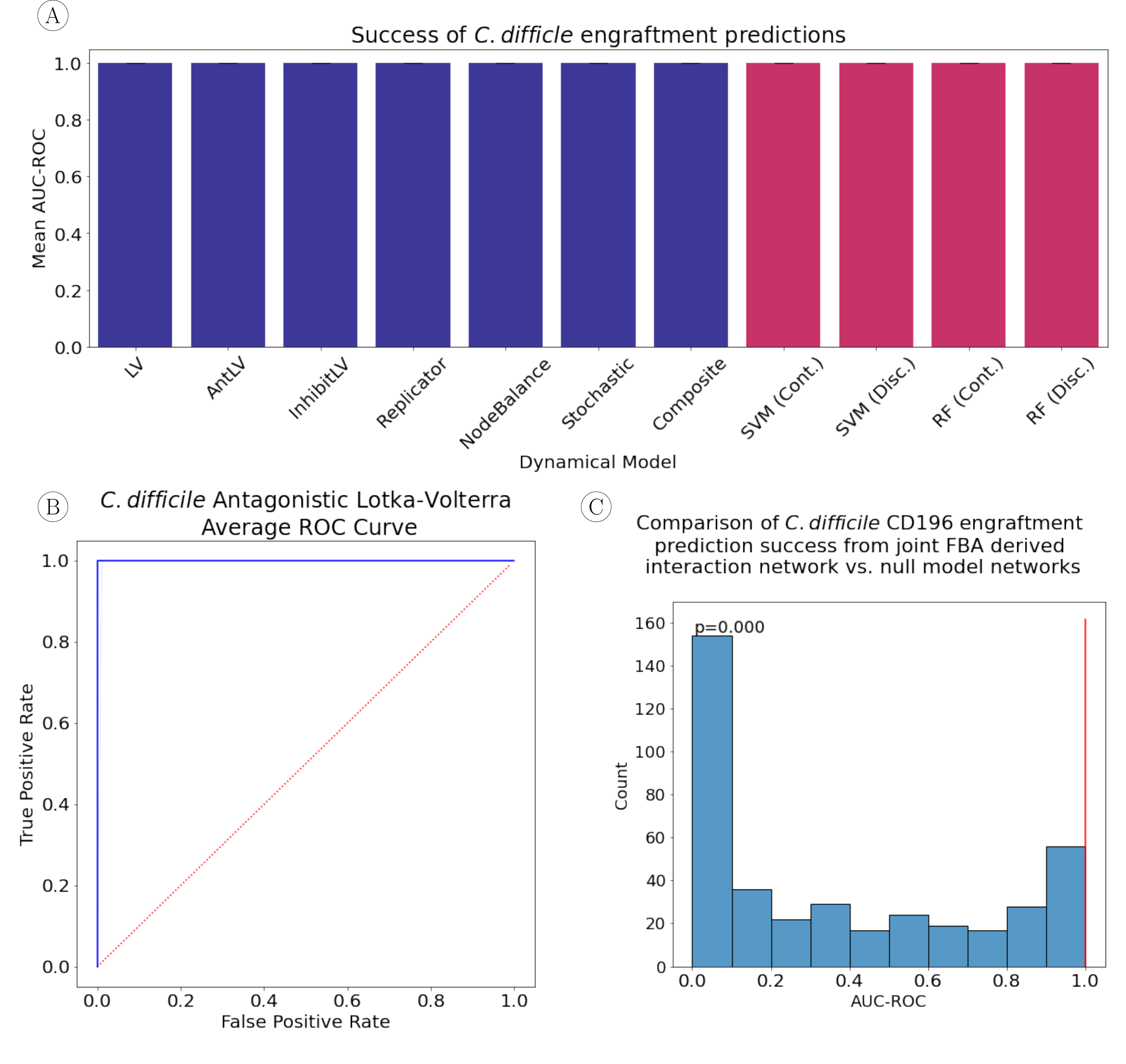}
    \caption{
    {\bf (A)} All network dynamics (blue bars) tested for the four AGORA \emph{C. difficile} models and the standard machine learning techniques (red bars) accurately predicted the engraftment versus non-engraftment as determined by the mean area under the receiver operator characteristic curve (AUC-ROC) (with 1 being perfect classification of engraftment and 0.5 being random chance). LV = Lotka-Volterra, AntLV = antagonistic Lotka-Volterra, inhibitLV = self-inhibition Lotka-Volterra, Replicator = replicator equation, NodeBalance = linear node balancing, Stochastic = random walk stochastic model, Composite = average of the previous six, SVM (Cont.) = support vector machine with continuous data, SVM (Disc.) =  support vector machine with discretized data, RF (Cont.) = random forest with continuous data, RF (Disc.) = random forest with discretized data. {\bf (B)} Example receiver operator characteristic curve (ROC) for a \emph{C. difficile} genome visualizing the perfect prediction of engraftment --- all true positives are identified before any false positives. {\bf (C)} Example comparison between the antagonistic Lotka-Volterra model (antLV) predictions of \emph{C. difficile} CD196 engraftment with the network parameters from joint FBA (red line) versus 400 null model networks (blue histogram). Null model networks consist of random interaction parameters generated using a label-swapping procedure to preserve the distribution of edge weights from the AGORA network. The clear superior performance of engraftment predicted with the AGORA network versus random networks indicates some level of biological significance.
    }
    \label{fig:REScdifficile}
\end{figure}

\begin{figure}
    \centering
    \includegraphics[scale = 0.3]{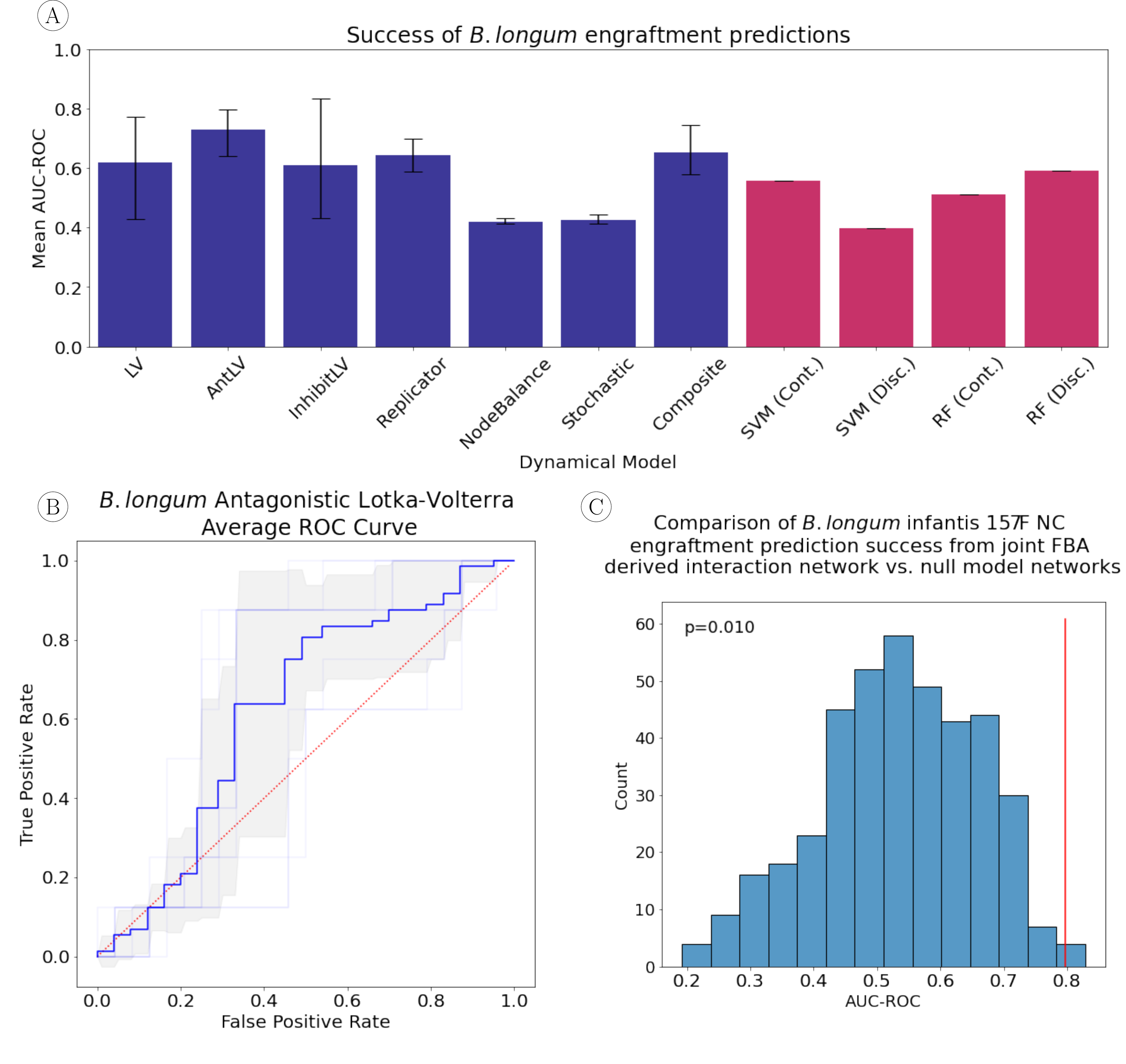}
    \caption{
       {\bf (A)} The power to predict engraftment versus non-engraftment of the nine AGORA \emph{B. longum} models varied across network dynamics (blue bars) and the standard machine learning techniques (red bars) as determined by the mean area under the receiver operator characteristic curve (AUC-ROC) (with 1 being perfect classification of engraftment and 0.5 being random chance). LV = Lotka-Volterra, AntLV = antagonistic Lotka-Volterra, inhibitLV = self-inhibition Lotka-Volterra, Replicator = replicator equation, NodeBalance = linear node balancing, Stochastic = random walk stochastic model, Composite = average of the previous six, SVM (Cont.) = support vector machine with continuous data, SVM (Disc.) =  support vector machine with discretized data, RF (Cont.) = random forest with continuous data, RF (Disc.) = random forest with discretized data. {\bf (B)} Average receiver operator characteristic curve (ROC) for \emph{B. longum} genomes visualizing the positive power of antagonistic Lotka-Volterra dynamics to predict engraftment --- true positives are identified before false positives (the curve is above the diagonal). {\bf (C)} Example comparison between the antagonistic Lotka-Volterra model (antLV) predictions of \emph{B. longum} infantis 157F NC engraftment with the network parameters from joint FBA (red line) versus 400 null model networks (blue histogram). Null model networks consist of random interaction parameters generated using a label-swapping procedure to preserve the distribution of edge weights from the AGORA network. The clear superior performance of engraftment predicted with the AGORA network versus random networks indicates some level of biological significance.
    }
    \label{fig:RESblongum}
\end{figure}

\begin{figure}
    \centering
    \includegraphics[scale = 0.3]{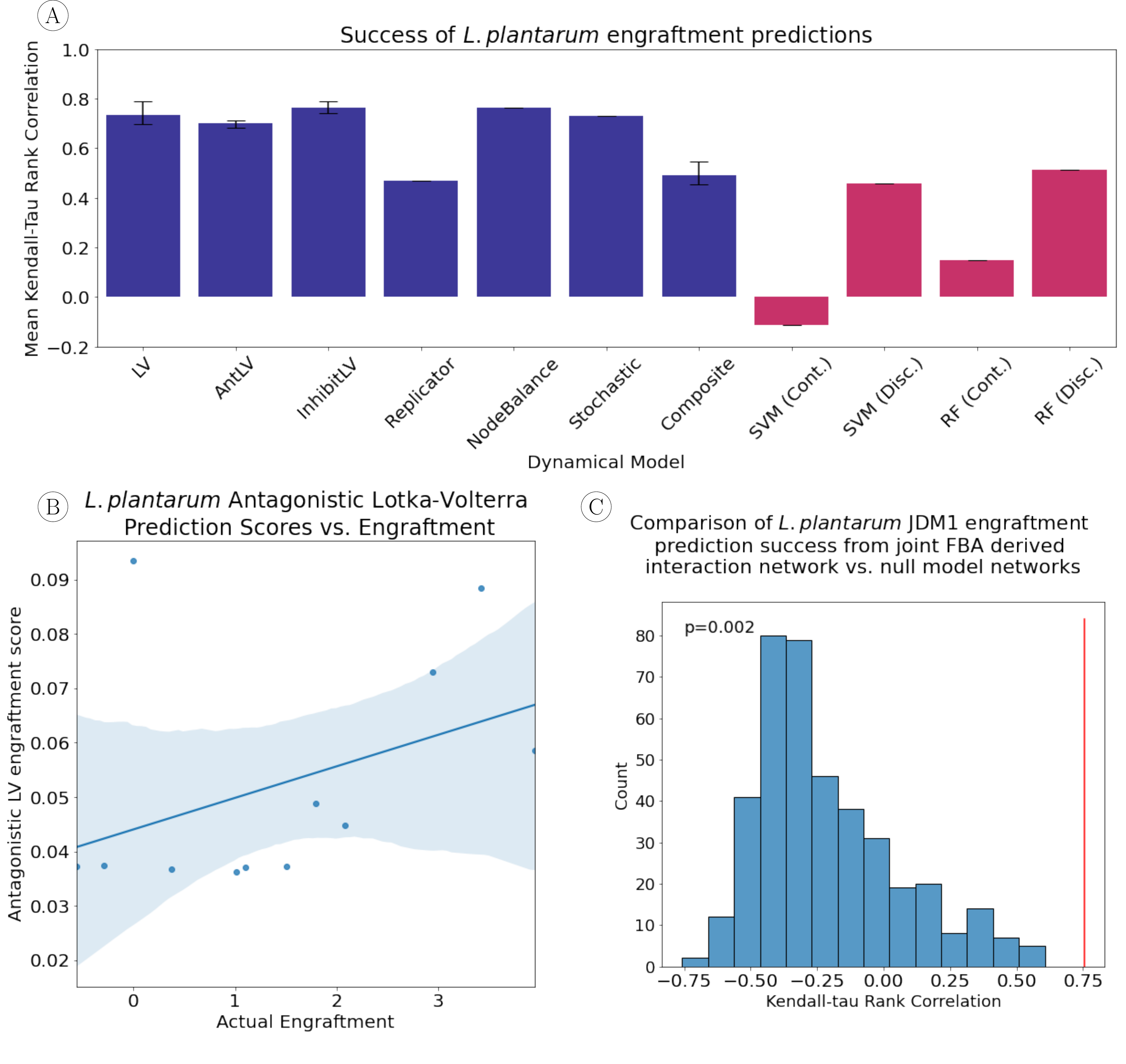}
    \caption{
          {\bf (A)} The power to predict engraftment versus non-engraftment of the three AGORA \emph{L. plantarum} models varied across network dynamics (blue bars) and the standard machine learning techniques (red bars) as determined by Kendall-tau rank correlation between sample score and measured engraftment (with 1 being perfect classification of engraftment and 0 being random chance).  We use Kendall-tau rank correlation as a metric of classifier success for the \emph{L. plantarum} because binary classification was unavailable for this data-set. Instead, measured engraftment is measured as the log-ratio of final time-point to initial time-point relative abundance. LV = Lotka-Volterra, AntLV = antagonistic Lotka-Volterra, inhibitLV = self-inhibition Lotka-Volterra, Replicator = replicator equation, NodeBalance = linear node balancing, Stochastic = random walk stochastic model, Composite = average of the previous six, SVM (Cont.) = support vector machine with continuous data, SVM (Disc.) =  support vector machine with discretized data, RF (Cont.) = random forest with continuous data, RF (Disc.) = random forest with discretized data. {\bf (B)} Average test-score vs. measured engraftment with linear regression best-fit line and $95\%$ confidence interval visualizing the positive power of antagonistic Lotka-Volterra dynamics to predict engraftment -- the slope of the regression line is positive. {\bf (C)} Example comparison between the antagonistic Lotka-Volterra model (antLV) predictions of \emph{L. plantarum} JDM1 engraftment with the network parameters from joint FBA (red line) versus 400 null model networks (blue histogram). Null model networks consist of random interaction parameters generated using a label-swapping procedure to preserve the distribution of edge weights from the AGORA network. The clear superior performance of engraftment predicted with the AGORA network versus random networks indicates some level of biological significance.
    }
    \label{fig:RESlplantarum}
\end{figure}

In \cref{fig:REScdifficile,fig:RESblongum,fig:RESlplantarum}, we report the AUC-ROC of each dynamical model for each experiment, as well as the AUC-ROC for support vector machine and random forest classification or regression (see \cref{fig:models} for the dynamics used). Tables of these results and associated estimated p-values can be found in the supplementary file \verb|supp_tables.pdf|, (results in Tables 1,3 and 6, p-values in Tables 2,4,7, and standard machine learning's performance in Tables 5 and 8). Study data from \citep{battaglioli2018clostridioides} displayed clear differences between engrafter and non-engrafter samples in both species composition and $\alpha$-diversity (see Supplementary Figure 3 in \verb|supp_figures.pdf|). For this reason, any reasonable classifier should be expected to differentiate the two groups. Indeed, our method produced perfect classification on the \emph{C. difficile} data-set for every of choice of dynamics or GEM representing the invading \emph{C. difficile}. We may consider this performance a necessary, but not sufficient criteria for the quality of the method. 

On the other hand, data from the studies \citep{maldonado2016stable,huang2021candidate} was much more muddled. In both of these studies, our method generally outperforms the classical machine-learning techniques to which we compared it. The Lotka-Volterra dynamics performed the best, especially when modified to be made antagonistic or inhibitory. Inspection of individual simulations revealed that this improvement is due to those modifications preventing finite-time blow up of solutions. 

We tested three general types of community model that can be associated with a network of interactions: generalized Lotka-Volterra dynamics (gLV, antagonistic LV, and inhibitory LV), replicator dynamics, and linear dynamics (node balancing and stochastic). The two linear models we used were both based on balance within a network; one based on balancing flux through the network nodes and the other based on the equilibrium of a stochastic system. Neither model performed well, suggesting that the low complexity of linear modeling is not sufficient for prediction. We observe much better predictive power from the two types of quadratic models used. The main difference between the generalized Lotka-Volterra model and the replicator model is that the replicator model includes a ``community effect" that is equal for every member of the community. Although this effect has been shown to be useful in predicting community composition \citep{gjini2021ratio,madec2020predicting}, it does not improve predictive power of our method on the data that we tested.


\subsection{Interpreting Mechanism - Sensitivity to Hyper-parameters}

One major advantage to any mechanistic model is that we may measure the sensitivity of our results to perturbations in the model. Here, we investigate the results of perturbing the model of the \emph{B. longum} experiments in two ways - simulating ``knock-out" experiments and computing sensitivity to changes in interaction strength. 

First, we simulate knock-out experiments by removing a taxa from the full AGORA network. As a result, this organism will be removed from any sample it was previously present in. We then measure the effect this has on our prediction of \emph{B. longum}'s likelihood of engrafting based on the antagonistic Lotka-Volterra dynamics.

Next, we measure the sensitivity of \emph{B. longum} growth to each interaction parameter in the antagonistic Lotka-Volterra model. That is, we measure the effect of perturbing each parameter individually on the simulated abundance of \emph{B. longum} at equilibrium according to antagonistic Lotka-Volterra dynamics.

\subsubsection{Sensitivity to Community Members}

In order to investigate how sensitive \emph{B. longum} engraftment is to each of the other species present in any sample of the \emph{B. longum} experimental data set, we simulate ``knock-out" experiments and observe the effect this has on our prediction of engraftment. We used the model of \emph{B. longum} infantis 157F NC to model invading \emph{B. longum} because this choice provided the best predictive value when using the full AGORA network and antagonistic Lotka-Volterra dynamics (see Table 3 of \verb|supp_tables.tex|). For each simulated knock-out, we removed a taxa from every sample it was present in and repeated our predictive procedure, using only antagonistic Lotka-Volterra dynamics for our prediction. We recorded the difference in sample score for engraftment of \emph{B. longum} (\cref{score}) as well as the change in AUC-ROC of our set of predictions.

We present the five knock-outs that had the largest average effect on our sample score across all samples of the \emph{B. longum} data set in \cref{tab:kos}. It is interesting to note the difference that each knock-out has on the predictive power of the model. For instance, removing \textit{L. delbrueckii} subsp. bulgaricus ATCC BAA 365, which on average increases the likelihood of predicting that \emph{B. longum} will engraft, causes our model to have very poor predictive value (AUC-ROC of $0.40$). This suggests that detection of \textit{L. delbrueckii} is important in prediction. Interestingly, this effect is due to a difference in knock-out effects between actual engrafters and actual non-engrafters. On average, \textit{L. delbrueckii} subsp bulgaricus ATCC BAA 365 knock-out increased the engraftment score of \emph{B. longum} by $0.086$ in non-engrafting samples and only $0.02$ in engrafting samples.

This experiment demonstrates that incorporating mechanism into prediction can provide useful insight beyond prediction. Because our model considers the network of interactions between microbial taxa, we are able to experiment with the effects of each individual taxa by using simulated experiments like knock-outs.

\begin{table}
    \centering
\begin{tabular}{llll}
\toprule
{} & \parbox[t]{2cm}{Sample \\Proportion} &  \parbox[t]{2cm}{Avg Score \\Difference} &  \parbox[t]{2cm}{AUC-ROC \\Difference} \\
\midrule
\parbox[t]{5cm}{\textit{L. delbrueckii} \\subsp bulgaricus ATCC BAA 365} &         0.46875 &              0.069928 &          -0.395833 \\
\textit{B. crossotus} DSM 2876                    &         0.84375 &             -0.016823 &          -0.114583 \\
\parbox[t]{5cm}{\textit{L. delbrueckii} \\subsp bulgaricus ATCC 11842} &         0.46875 &             -0.015429 &           0.020833 \\
\textit{M. timidum} ATCC 33093                   &         0.06250 &              0.014608 &           0.020833 \\
\textit{G. pamelaeae} 7 10 1 bT DSM 19378        &         0.81250 &              0.009144 &          -0.072917 \\
\bottomrule
\end{tabular}
    \caption{Top 5 most impactful knock-out experiments by change in sample score (Avg. Score Difference). The ``Sample Proportion" column gives the proportion of samples in the data-set that contain the organism that was knocked out. The final column shows the impact on our predictions of removing the organism from the analysis.}
    \label{tab:kos}
\end{table}


\subsubsection{Sensitivity to Interactions}

Our method is based on the network of interactions between microbial species, which ties together the individual interactions between each pair of species into one complex picture. This means that the interaction between two species unrelated to \emph{B. longum} may have an effect \emph{B. longum}'s growth. We can compute this effect by computing how \emph{B. longum}'s simulated abundance changes as we vary each parameter. Precisely, we can compute the derivative of \emph{B. longum} with respect to each parameter in the model. We once again limit this analysis to \emph{B. longum} infantis 157F NC.

We compute the sensitivity of \emph{B. longum}'s growth according to antagonistic Lotka-Volterra dynamics to each interaction parameter directly using the chain rule (see, for example \citep{zi2011sensitivity}). This has the following form:
\begin{equation}\label{paramdiff}
    \frac{\partial}{\partial t} \left(\frac{\partial x_i}{\partial \beta_{kl}}\right) = x_ix_k\delta_{i = l} + x_i\sum_{j\neq i} \beta_{ji}\frac{\partial x_j}{\partial \beta_{kl}} + \left(1 + 2\beta_{ii} x_i + \sum_{j\neq i}\beta_{ji}x_j\right)\frac{\partial x_i}{\partial \beta_{kl}}.
\end{equation}
\Cref{paramdiff} allows us to solve a system of differential equations for each parameter in the network to determine the sensitivity of invader growth to that parameter. 

We present the results of this analysis, averaged across each sample in the data set, in \cref{tab:deriv}. It is unsurprising that the growth of \emph{B. longum} was most sensitive to two direct regulations in the network, i.e. regulations that have \emph{B. longum} as their target. However, we observe that \emph{B. longum} is also sensitive to the regulation of \emph{G. pamelaeae} by \emph{B. crossotus}, as well as a number of other regulations that are mediated by \emph{B. crossotus}. These indirect effects suggests that \emph{B. crossotus} plays a large role in \emph{B. longum} growth even if it does not directly regulate \emph{B. longum}. 

This analysis demonstrates one advantage to network based prediction. Networks synthesize sets of individual interactions into a cohesive whole, and in doing so reveal indirect interactions that may be crucial to predicting microbial community composition.   

\begin{table}
    \centering
\begin{tabular}{llr}
\toprule
                                     Source &                                      Target &  Avg. Sensitivity \\
\midrule
            \textit{B. crossotus} DSM 2876 &     \textit{B. longum} infantis 157F NC &                       0.115290 \\
\textit{G. pamelaeae} 7 10 1 bT DSM 19378 &     \textit{B. longum} infantis 157F NC &                       0.023920 \\
            \textit{B. crossotus} DSM 2876 & \textit{G. pamelaeae} 7 10 1 bT DSM 19378 &                      -0.015566 \\
           \textit{L. gasseri} ATCC 33323 &     \textit{B. longum} infantis 157F NC &                       0.015427 \\
            \textit{B. crossotus} DSM 2876 &                 \textit{D. invisus} DSM 15470 &                      -0.013609 \\
            \textit{B. crossotus} DSM 2876 &                   \textit{R. bromii} L2 63 &                      -0.010666 \\
            \textit{B. crossotus} DSM 2876 &                  Ruminococcus sp 5 1 39BFAA &                      -0.009550 \\
      \textit{M. smithii} ATCC 35061 &     \textit{B. longum} infantis 157F NC &                       0.004876 \\
            \textit{B. crossotus} DSM 2876 &       \textit{M. smithii} ATCC 35061 &                      -0.004033 \\
\textit{G. pamelaeae} 7 10 1 bT DSM 19378 &                  Ruminococcus sp 5 1 39BFAA &                      -0.003314 \\
\bottomrule
\end{tabular}

    \caption{Sensitivity of \emph{B. longum} growth to the 10 most impactful interaction parameters in the AGORA network (excluding self-loops).}
    \label{tab:deriv}
\end{table}

\section{Discussion}


Our work exams the scenario where the goal of probiotic intervention is a long-lasting alteration of the resident host microbiota. Given that the links between microbes and health have a basis in observations of the resident human microbiota \citep{gupta2020predictive}, one common assumption is that alterations to that resident microbiota would impact health outcomes. Along those lines, we demonstrate that genome-scale metabolic modeling can be used in a simple way to predict the outcome of species invasion experiments with minimal study data required. This suggests that genome-scale models (GEMs) can provide value in understanding microbial community dynamics from cross-sectional microbiome relative abundance profiling. 

Genome-scale metabolic modeling uses genomic data to predict the growth and resource use of a microbial population by considering the entire internal metabolism of the species. This technique can be extended to community modeling in a number of ways \citep{zomorrodi2012optcom,chan2017steadycom,diener2020micom,kim2022resource}, all with relatively high levels of complexity. Community dynamics can be inferred from GEMs using a model known as dynamic flux balance analysis (FBA), which uses a GEM for each taxa in the community to infer growth rates and dynamic resource usage \citep{brunner2020minimizing}. While dynamic FBA provides a complete picture of community dynamics according to GEMs, it represents a very complex model that is difficult to analyze and simulate, and still contains a set of unknown parameters. Here, we simplify the modeling considerably by assuming that these underlying dynamics lead to a network of emergent interactions between microbes, which we can infer from pairwise genome-scale modeling.

The generalized Lotka-Volterra model and other similar microbial network models are popular tools for understanding community dynamics \citep{gore2017,fisher2014,angulo2019theoretical}. Such models are difficult to parameterize for a myriad of reasons. Good parameterization requires relatively dense time-longitudinal data of absolute, rather than relative, community abundance \citep{bucci2016mdsine,kuntal2019web}. By taking advantage of genome-scale metabolic models (GEMs), we provide a parameterized network that does not require any time-longitudinal data. In fact, our method as presented here makes use of previously published GEMs, meaning that only binary presence/absence information is necessary for a prediction. Additionally, GEMs can be built directly from the genomes found in a sample using automated tools such as \emph{CarveMe} \citep{machado2018fast}, meaning that our method can be extended to include taxa not found in the AGORA database.

Another strength of our approach is it's interpretability. The {\it B. longum} engraftment analysis provides us with an example motivation, which would be to identify other partner microbes that could improve the responsiveness to the engraftment of a target probiotic. In this instance, there is a very strong link between {\it L. delbrueckii} and {\it B. longum} engraftment in our model. This is consistent with co-culture studies that show {\it L. delbrueckii} and {\it B. longum} tolerate stressful conditions better together than when they are alone \citep{sanchez2013co}, suggesting some type of cooperation between the species. While we cannot know the mechanism of this cooperation from literature, our modeling suggests that it is at least reflected in part by metabolism.
Similarly intriguing is the cooperative link between {\it B. crossotus} and {\it B. longum} given the known cross-feeding interactions between many other butyrate-producing organisms \citep{falony2006cross,de2011cross,rios2015enhanced}---a metabolite that Butyrivibrio produce in large amounts \citep{cotta2006family}.

Lastly, we note that we have previously shown that species-species interaction modeling, i.e. models built from interactions between microbes, do not capture the complexities of microbial community dynamics that emerge as communities change in composition \citep{brunner2019metabolite}. Here, we mitigate this shortcoming by determining interaction parameters from pairwise models under specific metabolic conditions, providing limited environmental context to our method. However, it is unlikely that these interactions remain constant as the microbial community manipulates its environment. We conjecture that prediction can be improved by accounting for changes in microbial interaction as the environment changes. In upcoming work, we demonstrate that dynamic FBA implies that a microbial community behaves according to a discrete sequence of interaction networks over time. Incorporating this dynamic behavior may improve prediction with only a modest increase in model complexity, as long as this sequence of networks can be efficiently determined.

\section{Methods}\label{sec:meth}


\subsection{Inferring interaction parameters}

To make predictions without fitting any model to time series data, we use a network of interactions implied by a technique known as flux balance analysis (FBA). FBA is a technique used to predict growth rates of microbes using genome-scale information about their internal metabolisms \citep{lewis2012}. This technique requires a genome-scale metabolic model (GEM) which represents the set of metabolic pathways of a microbe, as well as information about the environmental metabolites available. An optimization problem is then solved to predict a growth rate of the microbe. Similarly, a technique known as ``joint flux balance analysis" \citep{zomorrodi2012optcom,chan2017steadycom,kim2022resource} can be used to estimate growth rates of pairs of microbes using a GEM from each microbe.

We use flux balance analysis with a ``western diet" environmental condition to simulate growth of the 818 taxa included in the AGORA database of GEMs \citep{magnusdottir2017generation}. Next, we use joint flux balance analysis to simulate the growth of every pair of these 818 microbes. We use joint FBA with a set of resource allocation constraints, which have been shown to improve predictions in pairwise experiments \citep{kim2022resource}. We construct our interaction network by taking a complete, weighted, directed graph with the 818 taxa as nodes and the log-ratio of simulated growth in pairs and alone as edge weights. That is, if simulated growth of species $i$ alone is $x_i$, and growth of species $i$ when coupled in a pair with species $j$ is $x_{ij}$, we weight the edge from species $j$ to species $i$ as 
\begin{equation}
    w^*_{ji}  = \log\left(\frac{x_{ij}}{x_i}\right)
\end{equation}
The log-ratio is used for simplicity and ease of computation. We also re-scale our network so that all edge weights are in the interval $[-1,1]$, meaning we take as our edge weights
\begin{equation}
    w_{ji} = \frac{w^*_{ji}}{\max_{l,k}(|w^*_{lk}|)}.
\end{equation}
This rescaling is done for computational convenience, and can be viewed as a time-rescaling of the network dynamics which does not effect our predictions.

We also trim the graph of the weaker edges to reduce the effect of spurious relationships between metabolic models. We tested our method using the strongest $N\%$ of edges in absolute value for $N$ ranging from $10$ to $100$. Predictions for \emph{C. difficile} and \emph{L. platarum} did not show any pattern of effect of $N$, while predictions for \emph{B. longum} improved or stayed level as $N$ was decreased to some threshold, roughly $N=25$, after which results became less consistent (see supplemental Figures 7-15 in \verb|supp_figures.pdf|). Here, we present results using $N=25$, meaning we keep only the strongest quartile of edges in absolute value. The result is a network of $818$ nodes and $167281$ edges, with a mean absolute edge weight of $0.515$ and variance in absolute edge weight of $0.0382$. We will refer to this final network as the ``AGORA network". 

\subsection{Creating induced sub-graphs}

In order to carry out our analysis using a network constructed from the AGORA database of GEMs \citep{magnusdottir2017generation}, it was necessary to match the taxa present in the data with GEMs in the database. The \emph{B. longum} study originally published in \citep{maldonado2016stable} made available metagenomic sequence data, while the other two studies made taxonomic profiles available. We created taxonomic profiles for the \emph{B. longum} study using \emph{Bracken} \citep{lu2017bracken}.

For each data-set, we matched taxonomic profiles to the set of AGORA models by querying the NCBI taxonomy database for nearest match taxa and comparing to the known taxonomy of the AGORA models. Only taxa with a species-level match to an AGORA model were included in our analysis. The result is that our analysis was performed using only a subset of the taxa in each sample. For the \emph{C. difficile} study published in \citep{battaglioli2018clostridioides}, we used an average of $36.2\%$ of the relative abundance of each sample; for the \emph{B. longum study} published in \citep{maldonado2016stable}, this was an average of $8.42\%$ of the relative abundance; and for the \emph{L. plantarum} study published in \citep{huang2021candidate}, this was $14.8\%$ of the relative abundance. 


\subsection{Assessing Receptivity}\label{sec:dynamics}

In order to assess how receptive a sub-graph is to an invading taxa, we use six different dynamical systems that can be associated with a pairwise graph, along with a composite score using all six. Each dynamical system allows us to simulate the community to equilibrium and score the performance of the invading taxa by simulated final abundance or time to extinction. The composite score can then be calculated simply as an average of the scores from the six dynamical systems. The dynamical systems fell into three categories - generalized Lotka-Volterra models, replicator equation-based models, and linear models.

\subsubsection{Generalized Lotka-Volterra Models}

The generalized Lotka-Volterra (gLV) model \citep{edelstein2005mathematical} of a community of $N$ species is written as follows:
\begin{equation}
    \frac{dx_i}{dt} = x_i \left( 1 + \sum_{j=1}^N \beta_{ji} x_j\right).
\end{equation}
Notice that we use a version of this model that assumes an intrinsic growth rate of $1$ for each taxa because fitting accurate intrinsic growth rates would require additional data, while we wish to assess the performance of our methods without the need for parameter fitting. We use three variations of the generalized Lotka-Volterra models, based on adjustment to the edge weights. These are:
\begin{enumerate}[label = (\alph*)]
    \item a model with the network edge weights as interaction parameters, $\beta_{ji} = w_{ji}$, which we refer to as ``Lotka-Volterra dynamics",
    \item a model with alll edge weights shifted to be negative, $\beta_{ji} = \nicefrac{(w_{ji} - 1)}{2}$, which we refer to as ``antagonistic Lotka-Volterra dynamics", and
    \item a model with added self-inhibition, $\beta_{ji} = w_{ji}$ for $i\neq j$ and $\beta_{ii} = -1$, which we refer to as ``inhibitory Lotka-Volterra dynamics".
\end{enumerate}
Models (b) and (c) are adjustments to the generalized Lotka-Volterra model that we implemented to prevent uncontrolled growth in simulation. In simulation, the standard version (a) exhibited blow-up of solutions which slowed computation and worsened prediction. Model (b) prevents blow-up by assuming that all interactions must be negative, or antagonistic, and adjusts all interactions so their relative effects are the same but absolute effects are negative. Model (c) prevents blow-up by using logistic growth for each individual microbe in isolation, rather than exponential growth in the standard gLV model.

The gLV model can display an array of behaviors, including multi-stability and chaos. For this reason, we use a Monte-Carlo sampling approach to generating predictions, taking repeated random draws of initial conditions and simulating forward. In our experiments, we use a number of draws equal to the number of nodes in the network. For each trial, we score the network based on the simulated relative abundance of the invader at some large time $T$. It is possible that the invading taxa either dominates the community (i.e. relative abundance approaches $1$) or becomes extinct in our simulation. We take this into account by setting a score as follows:
\begin{equation}\label{score}
    s_i = \frac{1}{3}\left(\frac{\text{time to extinction}}{T} + \text{relative abundance at time $T$} + \frac{\text{time to community dominance}}{T}\right)
\end{equation}
for each trial $i$ and averaging over all trials.

\subsubsection{The Replicator Model}

Recently, the replicator equation has been proposed as a model to study microbial community dynamics \citep{madec2020predicting,gjini2021ratio}. This equation takes the form
\begin{equation}
    \frac{dx_i}{dt} = x_i \left(\sum_{j=1}^N \beta_{ji} x_j - \b{x}^TB\b{x}\right)
\end{equation}
where $B$ is the matrix of $\beta_{ji}$. We take the network edges weights as interaction parameters, $\beta_{ji} = w_{ji}$ and refer to this model as ``replicator dynamics". As with the generalized Lotka-Volterra model, this model can display an array of complex behaviors, so we again use a Monte-Carlo sampling approach, scoring each sample according to \cref{score} and averaging over all samples.

\subsubsection{Linear Models}

Finally, we use two linear models that arise from considering the network as a mathematical object rather than considering the biological reality of the microbial community. We include these models for two reasons. The first is that they are fundamental to the network structure of the community, and the second is that they are much more computationally efficient and so convenient to use. 

The first of these models, which we refer to as ``node balancing dynamics" is based on the idea of balancing a flow of mass into and out of each node. Precisely, we rescale the network edge weights $w_{ji}$ to be in the interval $[0,1]$, taking
\[
\hat{w}_{ji} = \frac{w_{ji} + 1}{2}
\]
because the notion of node-balancing requires only positive interactions. Next, we compute the following graph Laplacian. Let $A$ be the matrix of rescaled edge weights $\hat{w}_{ji}$ so that species $j$ and $i$ are in the sample being scored. Then the Laplacian $L$ is defined as
\begin{equation}
    L = A^T - D
\end{equation}
where the diagonal matrix $D$ is the weighted degree matrix of the re-scaled networks, $D_{ii} = \sum_{j} \hat{w}_{ij}$. We then score the network based on the dynamical model
\begin{equation}
    \frac{d\b{x}}{dt} = L\b{x}.
\end{equation}
Letting $k$ be the index in the vector $\b{x}$ corresponding to the invading taxa, we can compute a score for the network using the dominant eigenvector (i.e. the eigenvector corresponding to the largest eigenvalue) of $L$. That is, if $\b{v}^d$ is the dominant eigenvector of $L$, we take $s = v^d_k$ as the score the network. 

The second linear model is very closely related to the first, and is based on the idea of a Markov process jumping between the nodes of the network and so we refer to this model as ``stochastic dynamics". Again, let $A$ be the matrix of re-scaled edge weights $\hat{w}_{ji}$. From this, we construct a stochastic matrix $\hat{A}$ by re-scaling so that $\hat{A}_{ji} =\frac{A_{ji}}{\sum_k A_{jk}}$ - the row sums of $\hat{A}$ are $1$. $\hat{A}$ now represents the transition matrix of a Markov jump process that moves between the nodes of the network, and we score the network according to the equilibrium probability that the process is in the node corresponding to the invading taxa. This can be calculated by computing the left eigenvectors of the matrix $\hat{A}$ and noting the property of stochastic matrices that there exists a left eigenvalue $\lambda = 1$ and any corresponding eigenvector is a stationary distribution for the process. Precisely, if $\b{v}^1$ is the left eigenvector corresponding to the eigenvalue $\lambda = 1$ and $k$ is the index of the invading taxa, we score the network $s = v_k^1$.

\begin{figure}
    \centering
    \includegraphics[scale = 0.7]{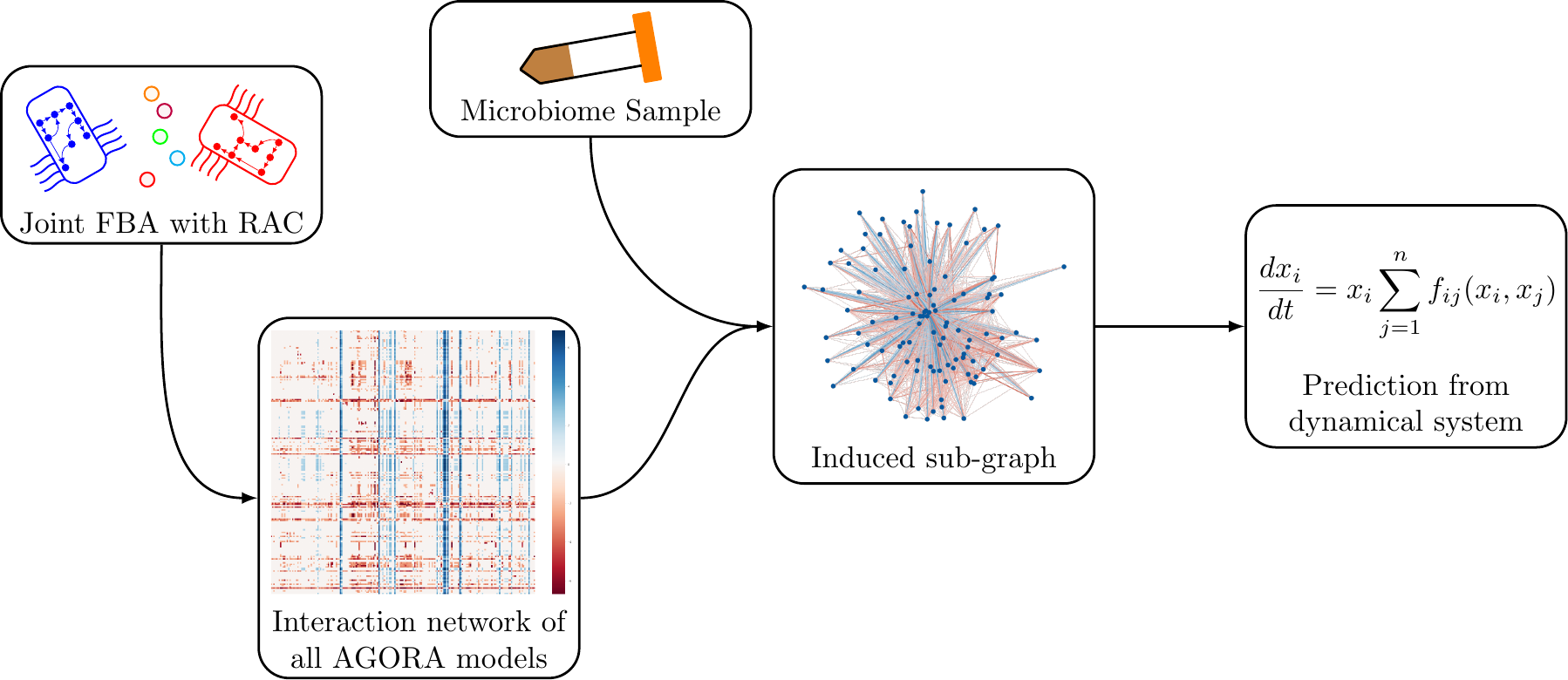}
    \caption{Schematic of the modeling process. In brief, we generate an interaction network of 808 AGORA models using pairwise joint flux balance analysis. To produce a prediction of engraftment for a given sample, we use the taxa present in the sample to generate an induced sub-graph of the full AGORA network. This is then used in a dynamical system to generate a prediction of engraftment.}
    \label{fig:schematic}
\end{figure}
 
\subsection{Prediction and evaluation}

We used data from \cite{battaglioli2018clostridioides,maldonado2016stable,huang2021candidate} in order to test the predictive power of the method. Data from \cite{battaglioli2018clostridioides} consisted of microbial communities taken from previously germ-free mice that had been inoculated with microbial communities from one of two dysbiotic or one of two healthy donors. In that study, these mice were then exposed to {\it Clostridioides difficile} to confirm that the dysbiotic mice were more susceptible to engraftment. Data from \cite{maldonado2016stable} consisted of bacterial community composition of fecal samples taken over the course of experiments in which {\it Bifidobacterium longum} AH1206 was administered as an oral probiotic to 23 subjects. Subjects were then differentiated into ``engrafters" and ``non-engrafters" based on the survival of the probiotic strain. Finally, data from \cite{huang2021candidate} consisted of bacterial community composition from fecal samples from 12 healthy participants who were administered {\it Lactiplanitbaccilus plantarum} HNU082 as an oral probiotic. In this study, we differentiated invasion success using the final time-point of each experiment.

Studies \cite{battaglioli2018clostridioides,maldonado2016stable} separated samples into successful and unsuccessful invasions, allowing us to use our method as a classifier. For these studies, we were able to compute scores for each sample, and vary the discrimination thresh-hold for the binary prediction across the observed values. From this, we could compute a ROC curve and integrate (commonly referred to as the ``area under the curve" of the ROC, or AUC-ROC). AUC-ROCs take values in the interval $[0,1]$ and, in general, an AUC-ROC greater than $0.5$ indicates positive predictive value of the model. We compared the AUC-ROC across the choices of dynamics and further compared to classification using support vector machine (SVM) and random forest (RF) classification \citep{scikit-learn}. The SVM and RF classifications were performed using both the relative abundances from the data as well as binary (presence/absence) forms of the data, which matches our method.

The study \cite{huang2021candidate} did not classify samples, and so we compared our predictions to the final time-point of that study using two well known measures of rank correlation: Kendall-tau rank correlation and Spearman rank correlation. Again, we compared across the choices of dynamics and to support vector machine regression and random forest regression \citep{scikit-learn}.

For all three studies, we estimated the significance of our predictions against predictions from a null-model created by replacing the AGORA network with a random network. To match the AGORA network in distribution, we used an approximate edge-swapping procedure. Precisely, we constructed a complete graph on the 818 AGORA models by drawing edge weights from the AGORA network, with replacement. We then filtered the network as in the construction of the AGORA network by keeping only the top $25\%$ of edges in absolute strength. We then computed AUC-ROC and rank correlations for predictions generated by this null model. We repeated this experiment 400 times in order to estimate a distribution of predictions from the null model.

Because we did not have strain specific models for the invading species in each data-set, we repeated our analysis for each model that matched at the species level with the invader.

\section{Acknowledgements}

The authors would like to thank Dr. Marie Kroeger for her thoughtful comments in helping to prepare this article.

J.B. was supported in this work by the  U.S. Department of Energy, Office of Science, Biological and Environmental Research Division using Award number F255LANL2018 and the Los Alamos National Laboratory Center for Nonlinear Studies.

\section{Data \& Code Availability}

Data from the three studies that we used to evaluate our method can be found at the following sources:
\begin{itemize}
    \item \citep{battaglioli2018clostridioides}: \url{https://www.ncbi.nlm.nih.gov/pmc/articles/PMC6537101/}
    \item \citep{maldonado2016stable}: \url{https://www.ncbi.nlm.nih.gov/bioproject/PRJNA324129/}
    \item \citep{huang2021candidate}: \url{https://microbiomejournal.biomedcentral.com/articles/10.1186/s40168-021-01102-0#Sec23}
\end{itemize}

Our method is implimented as the \emph{friendlyNets} package, available for download at \url{https://github.com/jdbrunner/friendlyNets} along with re-formatted data and python scripts for the analysis found in this paper.




\bibliography{probiotic_design}












\end{document}